\documentclass{nature}
\newcommand {\etal}{\textit{et al}.}
\newcommand {\EF}{$E_{\mathrm{F}}$}
\newcommand {\FS}{$F_{\mathrm{S}}$}
\newcommand {\vF}{$v_{\mathrm{F}}$}

\newcommand {\VG}{$V_{\mathrm{G}}$}
\newcommand {\CA}{Cd$_3$As$_2$}

\newcommand {\CZA}{(Cd$_{1-x}$Zn$_x$)$_3$As$_2$}

\newcommand {\SN}{Si$_3$N$_4$}
\newcommand {\AlO}{Al$_2$O$_3$}
\newcommand {\STO}{SrTiO$_3$}
\usepackage[dvipdfmx]{graphicx}
\usepackage{bm} 
\usepackage{pifont}
\usepackage{amsmath, amsthm, amssymb}
\usepackage{mediabb}
\usepackage[labelfont=bf]{caption}
\DeclareCaptionLabelSeparator{bar}{$|$\ }
\usepackage[labelsep=bar]{caption}
\usepackage{indentfirst}

\title{Quantized surface transport in topological Dirac semimetal films
}

\author{Shinichi Nishihaya$^{1}$, Masaki Uchida$^{1,2,*}$, Yusuke Nakazawa$^{1}$, Ryosuke Kurihara$^{3}$, Kazuto Akiba$^{3}$, Markus Kriener$^{4}$, Atsushi Miyake$^{3}$, Yasujiro Taguchi$^{4}$, Masashi Tokunaga$^{3}$ \& Masashi Kawasaki$^{1,4}$}

\begin{document}

\maketitle
\renewcommand{\baselinestretch}{1.5}\normalsize
\begin{affiliations}
 \item Department of Applied Physics and Quantum-Phase Electronics Center (QPEC), University of Tokyo, Tokyo 113-8656, Japan
 \item Precursory Research for Embryonic Science and Technology (PRESTO), Japan Science and Technology Agency (JST), Tokyo 102-0076, Japan
 \item Institute of Solid State Physics (ISSP), University of Tokyo, Kashiwa 277-8581, Japan
 \item RIKEN Center for Emergent Matter Science (CEMS), Wako 351-0198, Japan
\end{affiliations}

\renewcommand{\baselinestretch}{2}\normalsize
\begin{abstract}
Unconventional surface states protected by non-trivial bulk orders are sources of various exotic quantum transport in topological materials. One prominent example is the unique magnetic orbit, so-called Weyl orbit, in topological semimetals where two spatially separated surface Fermi-arcs are interconnected across the bulk. The recent observation of quantum Hall states in Dirac semimetal {\CA} bulks have drawn attention to the novel quantization phenomena possibly evolving from the Weyl orbit. Here we report surface quantum oscillation and its evolution into quantum Hall states in {\CA} thin film samples, where bulk dimensionality, Fermi energy, and band topology are systematically controlled. We reveal essential involvement of bulk states in the quantized surface transport and the resultant quantum Hall degeneracy depending on the bulk occupation. Our demonstration of surface transport controlled in film samples also paves a way for engineering Fermi-arc-mediated transport in topological semimetals.
\end{abstract}


Elucidating unconventional properties of surface/edge states protected by a non-trivial bulk order in topological materials has been one of the most important subjects in recent condensed matter physics\cite{topo1,topo2,topo3}. Being gapless types of topological materials, Dirac and Weyl semimetals (DSM/WSM) show unique open-arc surface states (Fermi-arcs) as a direct consequence of the monopole charges $\pm 1$ carried by the paired band crossing points (Weyl points)\cite{topo3,arc_ori} (Fig. 1a). One crucial feature of the Fermi-arcs is that, around their termination at the Weyl points, the wave functions are no longer localized at the surface but spatially extended to the inside, merging with the bulk states\cite{topo3}. Under a magnetic field, this unique connectivity enables the formation of an exotic orbit (Weyl orbit), where two Fermi-arcs on opposite surfaces are interconnected by bulk chiral modes ($N = 0$ Landau levels) transferring electrons across the slab thickness\cite{arc_tra_the1} (Fig. 1b), leading to the experimental detection of quantum oscillations\cite{arc_tra1,arc_tra2,arc_tra3}.

The recent theoretical prediction of Weyl orbit quantization\cite{arc_tra_the3,arc_tra_the4} and the following experimental observation of quantum Hall (QH) states in bulk nano-plates of the DSM {\CA}\cite{arc_tra2, arc_tra4, arc_tra5} has provoked further interest in the novel quantization physics in three-dimensional (3D) topological semimetals. The actual underlying mechanism for the quantized transport, however, has remained elusive and controversial. In a DSM consisting of two copies of WSM overlapping with opposite chirality, possible formation of topological-insulator-like surface orbits localized on each sample surface (Fig. 1c) has also been discussed\cite{arc_tra_the2, arc_tra5}. As originally predicted for quantum oscillations\cite{arc_tra_the1}, the existence of an additional phase term resulting from the bulk tunnelling process can be a clear indication of Weyl orbit transport, though it is still questionable and not theoretically verified whether such a phase term originating from the bulk state can be even applied to quantum Hall effect where exact quantization of electron phases within a two-dimensional (2D) gapped energy structure is necessary. In this context, it is highly meaningful to adopt thin film samples with excellent tunability of dimensionality, Fermi level ({\EF}), and interfaces to investigate the bulk contribution to the surface QH states in DSM.
Here, we report 2D quantum oscillations and QH states observed in 3D DSM {\CA} films. By controlling the topological phase transition with Zn doping, we have clarified that the quantized conduction originates from the surface state of the DSM phase. The possible confusion between the surface QH state and the conventional QH state induced by confined bulk states\cite{CA_kwsk1} is avoided by designing an appropriate film thickness. Through careful analysis of the surface QH states  particularly with electrostatic control of {\EF}, we reveal the essential involvement of the bulk states and the QH degeneracy highly dependent on the bulk occupation.


Magnetoresistances (MRs) of {\CZA} (112) oriented films were measured for different current $I$ and magnetic field $B$ configurations. First, we present in Fig. 1d out-of-plane transverse ($B \perp I$, $\theta = 90^{\circ}$) and longitudinal MRs ($B \parallel I$, $\theta = 0^{\circ}$) of a 100 nm thick {\CA} film measured with a pulsed field up to 55 T. Focusing on the quantum oscillations, the longitudinal MR is characterized by a single oscillation frequency corresponding to the 3D bulk Fermi surface. In the transverse MR, on the other hand, another oscillation component with a higher frequency (a larger Fermi surface area {\FS}) is superimposed on that of the bulk state (see Supplementary Fig. 4 and Supplementary Note 3 for Fourier analyses). Especially around the valleys of the bulk oscillations in higher fields, there appear significant drops in the resistance $R_{xx}$ and quantized plateaus in the Hall resistance $R_{yx}$. The QH effect is a feature of a high-mobility 2D electron gas (2DEG), suggesting that the additional conduction component in the transverse MR is of 2D nature developed from the surface states.

Figures 1e and 1f show temperature dependence of the oscillatory components in MRs and the effective mass $m^{*}$ extracted by the conventional fitting method (see Supplementary Fig. 2) as a function of $1/B$. Up to 20 T, $m^{*}$ is around 0.06 $m_e$ ($m_e$ is the electron mass), in good agreement with previous reports on those of {\CA} bulks\cite{CA_tra1, CA_tra2, CA_tra3}. By contrast, $m^{*}$ becomes 5$\sim$6 times larger for the higher-frequency oscillations observed above 20 T. 
The quicker decay of the 2D quantum oscillations as compared to the bulk oscillations against temperature rise is consistent with the surface transport reported for (112) bulk nano-plates in previous studies\cite{arc_tra2, arc_tra3}. The difference in the underlying band dispersions between the bulk and the 2D transport can also be found in the indices of Landau levels (LLs). While the bulk state reaches $N = 2$ already at 30 T, the QH plateau indicates much larger filling factors $\nu$ such as $\nu = 15$, 13, and 12 which cannot be explained by the bulk LLs.


To gain a more direct insight, we next present how the 2D transport changes through the chemical-doping-induced topological phase transition as summarized in Fig. 2. Figures 2b-2d show the MRs of the Zn-doped samples ($x$ = 0.14, 0.17, and 0.21) (see also Supplementary Fig. 3 for lower-$x$ data). To explicitly display the differences in the oscillation components in the transverse and longitudinal MRs, their second derivatives are also presented in Figs. 2e-2g. For $x = 0.14$ and 0.17, an additional oscillation component of a higher frequency coexists with the bulk oscillations in the transverse MR similarly to the non-doped case presented in Fig. 1d (see also Supplementary Fig. 4). For $x = 0.21$, on the other hand, only a single oscillation component dominates, exhibiting no sign for the existence of an additional Fermi surface. The chemical doping of Zn reduces the spin-orbit coupling and lifts the band inversion in {\CA}, inducing the topological phase transition from DSM to a trivial insulator\cite{CZA1,CZA2,CZA3} (Fig. 2a). The absence of the 2D transport in the more heavily doped sample indicates that its origin lies in the surface Fermi-arcs of the DSM phase which disappears as the system crosses the topological phase transition. 
The progressive shrink of the surface oscillation frequency upon Zn doping is also consistent with this viewpoint (Supplementary Fig. 5).
We note that for all samples with $x \leq 0.21$ the mean free path is larger than the film thickness (see Supplementary Fig. 1), fulfilling the requirement for observing the Weyl orbit state\cite{arc_tra_the1}. It is also worth mentioning that the disappearance of the surface transport coincides with the suppression of the negative longitudinal MR ascribed to the chiral charge pumping in the less doped region ($x < 0.21$)\cite{CZA3}, implying their common origin lying in the topological nature.


Since {\CA} has a small effective mass, a slight confinement (thickness $t <$ 40 nm) can easily induce sub-band formation in the bulk state, leading to 2D bulk QH states\cite{CA_kwsk1}. Similar observations were later reported in molecular beam epitaxy grown {\CA} films ($t \leq$ 35 nm)\cite{CA_MBE} and Zn doped films ($t \leq$ 35 nm)\cite{CA_kwsk2}. However, we emphasize that the 2D transport observed here is qualitatively different from such a confinement-induced one in that it is accompanied with a 3D bulk state. To clarify this point, we compare in Fig. 3 two {\CZA} samples with similar sheet carrier density but different film thicknesses (35 nm and 95 nm). For the thinner sample, as indicated by the absence of quantum oscillations in the longitudinal MR, the bulk state is confined into the 2D sub-bands. These sub-bands hold a high-mobility 2DEG, giving rise to QH states with $R_{xx}$ at zero. The band properties such as effective mass $m^*$ and Fermi velocity {\vF} in the 2D sub-bands are almost identical to those of the original 3D bulk\cite{CA_kwsk1, CA_kwsk2}. For the thicker film, the bulk Fermi surface remains 3D, meaning that the bulk state is still in a gapless DSM phase. While the surface QH states appear at higher fields, the bulk quantum oscillations remain on the background (see also Supplementary Figs. 6 and 7 for detailed field angle dependence). The QH states develop especially around the bulk oscillation valleys where the density of states or the conduction contribution from the bulk state reaches a minimum.

The sheet carrier-density involved in a QH state can be calculated through the relation $n_{\mathrm{2D,QHE}} = \nu eB /h$ ($e$ is the elementary charge and $h$ the Planck constant). For the $x = 0.15$ sample in Fig. 3d, the estimated density of $n_{\mathrm{2D,QHE}}$ = 1.1 $\times 10^{12}$ cm$^{-2}$ matches up with the total sheet carrier density of the film obtained from $R_{yx}$ at low fields ($n_{\mathrm{2D,Hall}}$ = 1.1 $\times 10^{12}$ cm$^{-2}$). 
This indicates that all sheet carriers including the bulk ones are involved in the surface QH states, even though the bulk state is not completely depleted under the applied field range. The similar tendency has been commonly observed for the QH states in bulk nano-plate samples\cite{arc_tra4, arc_tra5}, though this fact has not been well appreciated so far (See also Supplementary Fig. 8 and Supplementary Note 6). The quantization takes place consistently at the field positions determined by the total sheet carrier density ($1/B = \nu e /n_{\mathrm{2D,Hall}}h$). Therefore, it seems not appropriate to simply apply the original Weyl orbit formula characterized by the thickness-dependent phase term\cite{arc_tra_the1}, to the quantized transport. The involvement of bulk carriers also indicates that the bulk and surface conductions are neither independent nor parallel any more in the QH states. This characteristic feature may reflect the unique nature of the magnetic orbits in DSM where the bulk and surface states are woven together through their connectivity at the Weyl points, as discussed later.


We next present the evolution of the surface QH states with tuning {\EF} to further demonstrate the bulk contribution to the surface transport. Figure 4 summarizes gate-voltage {\VG} scans of the MRs for a $x = 0.19$ sample at each field step. The $x = 0.19$ sample is still in the DSM phase, showing surface transport as indicated by the appearance of QH effect in Fig. 4a (see also Supplementary Fig. 6). In Figs. 4b and 4c, {\VG} and $B$ mappings of oscillation peaks and valleys of $R_{xx}$ are presented by taking the second derivative of the transverse and longitudinal MRs. With increasing the negative bias, a clear crossover from the bulk-dominant conduction into the surface QH effect appears around {\VG} $= -5$ V. The {\VG} scan of carrier density reveals the successive change in the filling factors $\nu$ of the surface QH states. In particular, a four-fold degeneracy is observed, for example in the {\VG} scan at 6 T. If the magnetic orbits are completed only within the surface states, the degeneracy of the QH states should be 2. Thus, the observed degeneracy of 4 suggests that the actual orbits are formed by involving the bulk states, where the additional degeneracy can be explained by two kinds of bulk states forming different orbits.

The four-fold degeneracy of QH states has been also observed in {\CA} bulk nano-plates ($t \sim$ 80 nm) especially around the field range before the bulk state reaches the quantum limit\cite{arc_tra2}. It has been suggested that two of the confinement-induced sub-bands derived from the bulk chiral mode ($N = 0$ LL) give rise to two pairs of Weyl orbits to explain the degeneracy. Our successive {\VG} and $B$ scans in Fig. 4, however, reveal only the Landau splitting of the bulk state and no clear signature of sub-band splitting. Therefore, the underlying bulk states for the four-fold degeneracy of the surface QHE are most likely the two different bulk Landau states. Indeed, in Fig. 4b, the four-fold degeneracy appears in the region where the bulk $N = 0$ and $N = 1$ levels remain occupied below {\EF}. Considering the non-chiral nature of the $N = 1$ level, the surface magnetic orbits mediated by this LL should be conventional ones connecting the double Fermi-arcs on each surface, not the Weyl orbit expected for the orbits via the bulk chiral mode. As the bulk $N = 1$ level is depleted by increasing {\VG} or $B$, only the orbits via $N = 0$ level remains, reducing the total degeneracy to 2.

The question of great interest here would be whether the orbit mediated by the bulk $N = 0$ level is the Weyl orbit or the conventional surface orbit. In Fig. 4b, the QH degeneracy further lifts from 2 to 1 with increasing {\VG} or $B$. The same degeneracy lifting is confirmed in the high- field regime for our film samples (Figs. 1d, 2b, and 3d), and also for bulk nano-plates in previous studies\cite{arc_tra4, arc_tra5}. While these observations seem to contradict the naive expectation that Weyl orbits are always doubly degenerate due to the preserved symmetry, there are several effects which can lead to a degeneracy lifting of the Weyl orbits, such as the time-reversal-symmetry-breaking-induced transition into WSM\cite{NB}, or the symmetry-breaking-induced transition of Weyl orbits into conventional surface orbits\cite{arc_tra_the1} (see also Supplementary Note 8 for details). Both effects are expected to become more evident in the high-field or low-density regime. In the film sample case, their degeneracy lifting effect can be even enhanced by asymmetric conditions between the top and bottom surfaces such as induced by gating or interfaces. Thus, the observed degeneracy lifting does not exclude the Weyl orbit scenario. To directly identify the existence of the unconventional orbit, one may need to investigate the thickness dependence of surface quantum oscillations (not the surface QH effect) and confirm the thickness dependent phase term as predicted in the original theory\cite{arc_tra_the1}. Regarding this point, one important implication from the carrier density analysis is that once the QH states develop, such a phase term may become obscured as the quantization occurs simply according to the total sheet carrier density of the sample (Supplementary Fig. 8). Further investigations are desired to further clarify the detailed nature of the surface orbits realized in the {\CA} system and their underlying quantization mechanism.


Through the control of dimensionality, {\EF}, and band topology in {\CZA} films, we have demonstrated the emergence of surface QH states and their characteristic QH degeneracy dependence on the bulk LL occupation. To independently control each surface by accumulation/depletion or designing heterointerfaces with ferromagnets or superconductors is an important topic for further engineering the surface transport\cite{WO1,WO2,WO4}. The interface control can also be useful to probe possible interplays between the Fermi-arcs and the bulk topological transport in DSM such as non-local charge pumping by chiral anomaly\cite{CA_CA1, CA_CA2}, spin Hall effect\cite{DSM_SHE}, node-separation-gradient-induced pseudo-magnetic field\cite{WSM_FF} and exotic superconductivity\cite{CA_SC1,CA_SC2} through their connectivity in the momentum space. In this respect, our demonstration of the surface QH states controlled in thin films provides an essential platform for investigating the novel quantized transport phenomenon.


\begin{methods}
\noindent \textbf{Film growth and device fabrication}

The epitaxial {\CZA} (112) films were grown on {\STO} (100) single-crystal substrates by the combination of pulsed laser deposition technique and subsequent thermal annealing\cite{CA_kwsk1,CZA3,CA_kwsk10}. The film thickness is designed to be around 85$\sim$100 nm to maintain the three-dimensionality of the bulk state. To strictly define the current path in magnetoresistance measurements, the films were deposited through a stencil metal mask and patterned into a Hall bar shape with channel width of 60 $\mu$m. After the deposition, they were capped by MgO (5 nm)/{\SN} (200 nm) to prevent their evaporation and oxidization in the subsequent thermal annealing at 600 $^{\circ}$C in air. For the electrostatic gating measurements, a top-gate configuration was fabricated by thinning the topmost {\SN} capping layer down to 10 nm by ion milling, and then depositing 40 nm thick {\AlO} as gate dielectric by atomic layer deposition, followed by the deposition of 50 nm thick Au as gate electrode (see also Supplementary Fig. 9 and Supplementary Note 7 for the device structure).

\noindent \textbf{Low-temperature magnetotransport measurements}

The magnetoresistance measurements up to 55 T were performed using a non-destructive pulsed magnet at the International MegaGauss Science Laboratory at the Institute for Solid State Physics of the University of Tokyo. Longitudinal resistance $R_{xx}$ and Hall resistance $R_{yx}$ were measured with either DC or AC excitation current, which was typically set at a few hundreds of $\mu$A to obtain a sufficiently good signal-to-noise ratio. While most of the measurements were conducted in a constant-current mode, we also adopted a constant-voltage mode for measuring the transverse MR of the low carrier density samples, where the resistance becomes significantly large at high fields. The transport measurements up to 9 T or 14 T were performed in Physical Property Measurement Systems (Quantum Design). Hall measurements with electrostatic gating were conducted using a lock-in technique, where the excitation current was kept constant at 0.5 $\mu$A with a frequency of 13 Hz. The contour mappings in Fig. 4 were obtained by scanning the gate voltage from 0 V by $-0.1$ V steps at each magnetic field.

\end{methods}

\newpage
\renewcommand{\baselinestretch}{1}
\renewcommand{\baselinestretch}{1.5}\normalsize

\begin{addendum}
 \item We thank M. Kawamura, H. Ishizuka, C. Zhang, H.-Z. Lu, N. Nagaosa, and Y. Tokura for fruitful discussions. This work was supported by JST PRESTO Grant No. JPMJPR18L2 and CREST Grant No. JPMJCR16F1, Japan, and by Grant-in-Aids for Scientific Research (B) No. JP18H01866 and (C) No. JP15K05140 from MEXT, Japan.
 \item[Author contribution] S.N., M.U. and M.Kawasaki designed the experiments. S.N., M.U., Y.N. synthesized the bulk targets with M.Kriener and performed thin film growth. S.N., M.U., R.K., K.A., A.M., and M.T. performed the high-field measurements. S.N. and M.U. analysed the data and wrote the manuscript with contributions from all authors. Y.T. and M.Kawasaki jointly discussed the results. M.U. and M.Kawasaki conceived the project. All authors have approved the final version of the manuscript.
 \item[Additional information] Supplementary information is available in the online version of the paper. Reprints and permissions information is available online at www.nature.com/reprints. Correspondence and requests for materials should be addressed to M.U. (email: uchida@ap.t.u-tokyo.ac.jp).
 \item[Competing interests] The authors declare that they have no
competing financial interests.
\end{addendum}

\newpage
\renewcommand{\baselinestretch}{1.3}\normalsize
\begin{figure}
\begin{center}
\includegraphics[width=16.6cm]{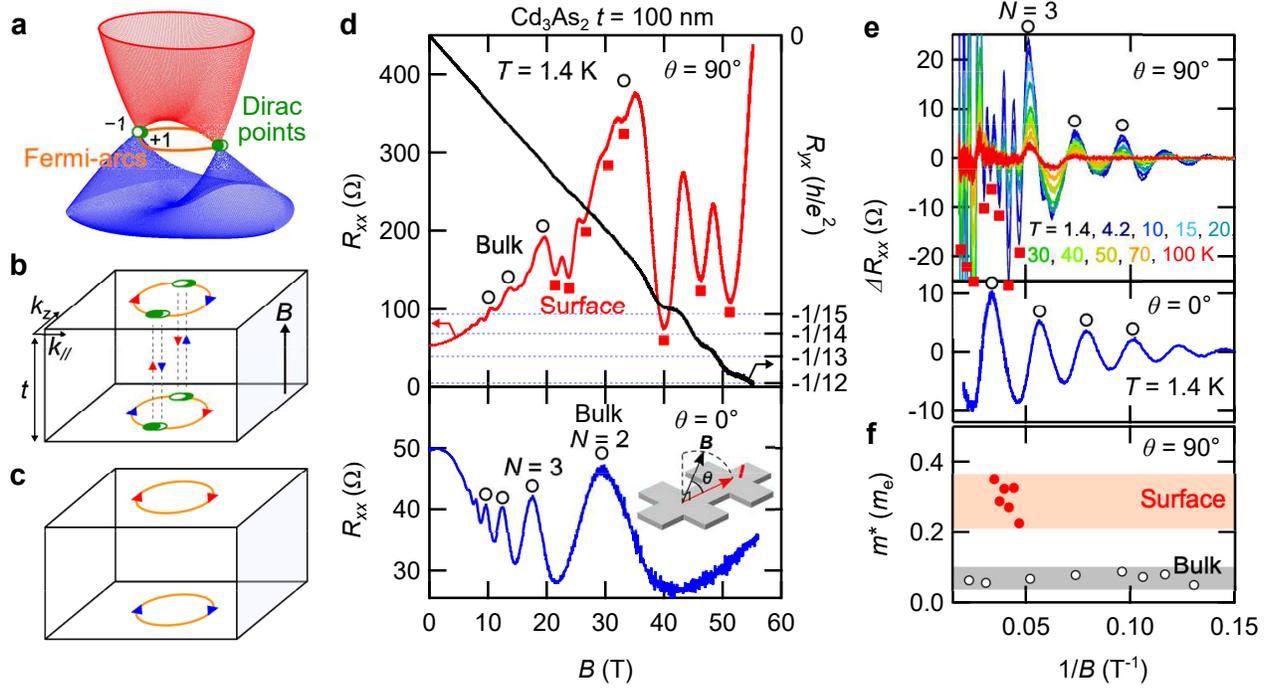}
\caption{
\textbf{Surface quantum transport in a {\CA} thin film.} \textbf{a}, Low-energy band dispersion of a Dirac semimetal (DSM). DSM consists of two copies of a Weyl semimetal superimposed with opposite chirality $\pm 1$. \textbf{b},\textbf{c}, Magnetic orbits mediated by the surface states in a DSM slab with thickness $t$. The two Fermi-arcs on the top and bottom surfaces are interconnected through the bulk chiral zero mode in \textbf{b}, while the two Fermi-arcs on the same surface deform into a closed Fermi pocket to form a conventional orbit localized at each surface in \textbf{c}. \textbf{d}, Out-of-plane transverse (upper panel) and longitudinal (lower panel) magnetoresistances (MRs) of a 100 nm thick {\CA} film. The right axis of the upper panel shows Hall resistance $R_{yx}$. $\theta$ indicates an angle between the field $B$ and current $I$. Filled squares denote the oscillation component observed only in the transverse MR, while open circles denote the 3D bulk oscillations observed in both the transverse and longitudinal MRs. Accompanied by sharp resistance drops in $R_{xx}$ in the higher field region, $R_{yx}$ shows plateau-like structures, indicating the appearance of 2D quantum Hall states. \textbf{e}, Temperature dependence of the oscillatory components in the transverse and longitudinal MRs and \textbf{f}, extracted effective mass $m^{*}$ plotted as a function of $1/B$. The quantum oscillations at higher fields are characterized by larger {\FS} and $m^{*}$ than the bulk one, which are ascribed to the surface state.
}
\end{center}
\end{figure}
\clearpage
\newpage

\begin{figure}
\begin{center}
\includegraphics[width=16.6cm]{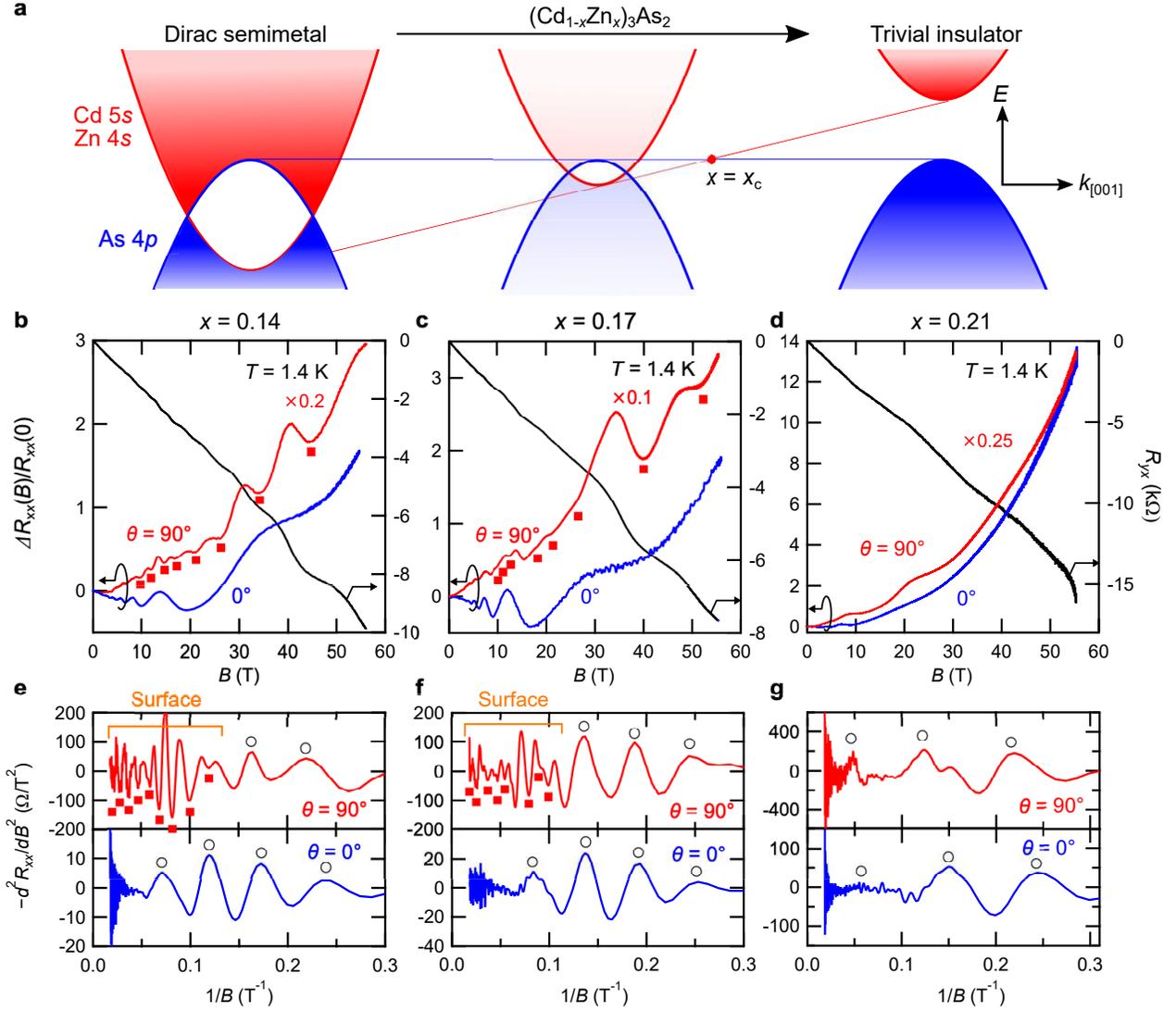}
\caption{
\textbf{Evolution of the surface transport through topological phase transition.} \textbf{a}, Zn-doping-induced topological phase transition from a Dirac semimetal (DSM) to a trivial insulator in {\CZA}. \textbf{b}-\textbf{d}, Transverse and longitudinal magnetoresistances (MRs) of {\CZA} films with different Zn concentrations ($x$ = 0.14, 0.17, and 0.21). \textbf{e}-\textbf{g}, Second derivative of the MRs plotted as a function of $1/B$. In addition to the bulk oscillations (open circles), an additional oscillation component (filled squares) is observed only for $x = 0.14$ and $0.17$, indicating the disappearance of the surface state for $x = 0.21$ which is above the critical composition $x_{c}$.
}
\end{center}
\end{figure}
\clearpage
\newpage

\begin{figure}
\begin{center}
\includegraphics[width=15.0cm]{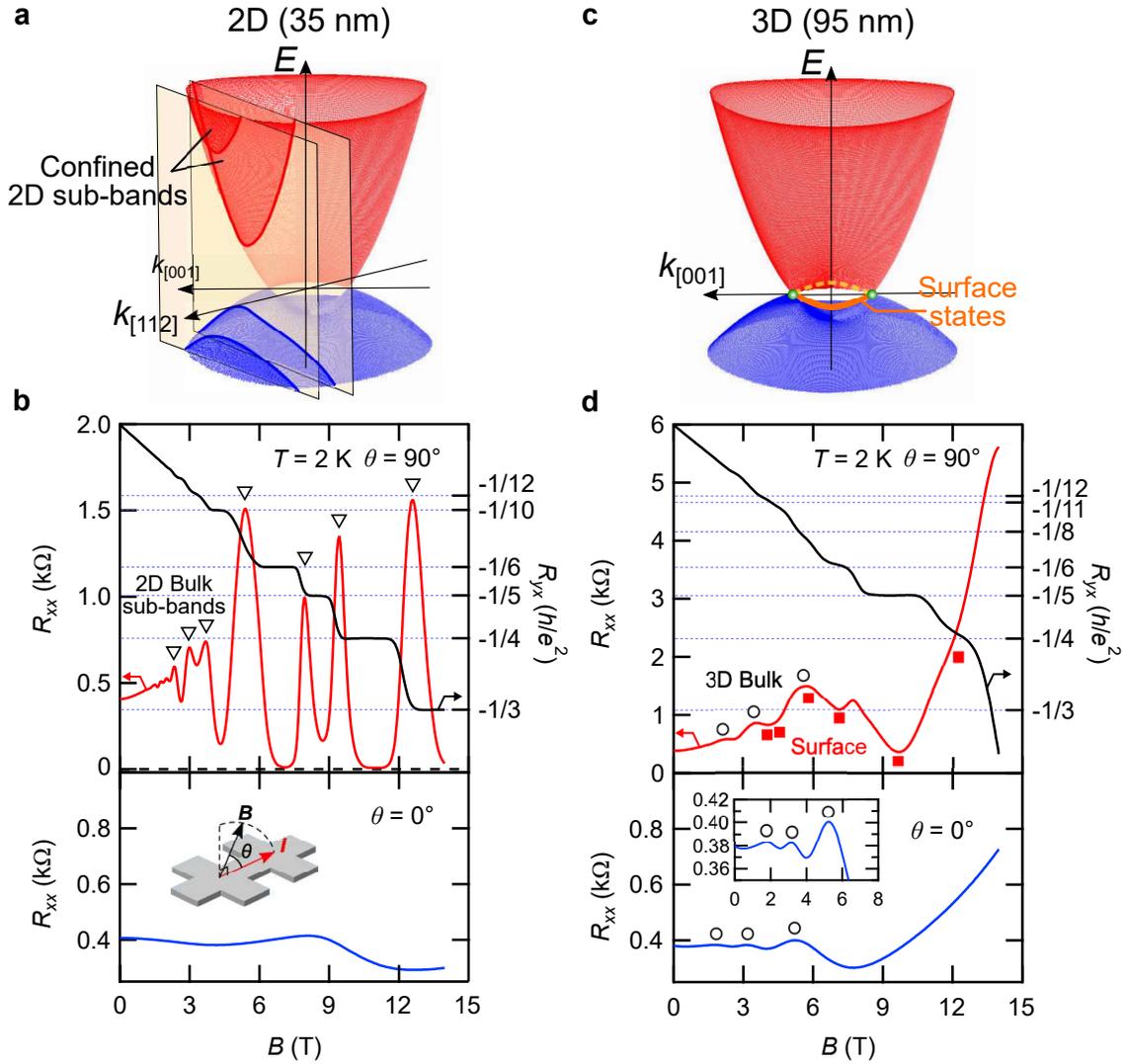}
\caption{
\textbf{Quantum Hall effect in 2D and 3D films.} \textbf{a}, Sub-band formation due to the quantum confinement effect in the [112] direction in a thinner film\cite{CA_kwsk1}. \textbf{b}, Quantum Hall (QH) states developed from the confined bulk states in a 35 nm thick {\CZA} film ($x$ = 0.13, $\theta = 90^{\circ}$, upper panel). The absence of quantum oscillations in the longitudinal magnetoresistance ($\theta = 0^{\circ}$, lower panel) indicates the 2D nature of the confined bulk Fermi surface. \textbf{c}, 3D Dirac semimetal phase with surface states in a thicker film. \textbf{d}, Weyl orbit QH states ($\theta = 90^{\circ}$, upper panel), and longitudinal magnetoresistance with clear quantum oscillations from the 3D bulk state ($\theta = 0^{\circ}$, lower panel) in a 95 nm thick film ($x$ = 0.15). The QH states develop against the background of bulk quantum oscillations, particularly around the bulk oscillation valleys.
}
\end{center}
\end{figure}
\clearpage
\newpage
\begin{figure}
\begin{center}
\includegraphics[width=16.6cm]{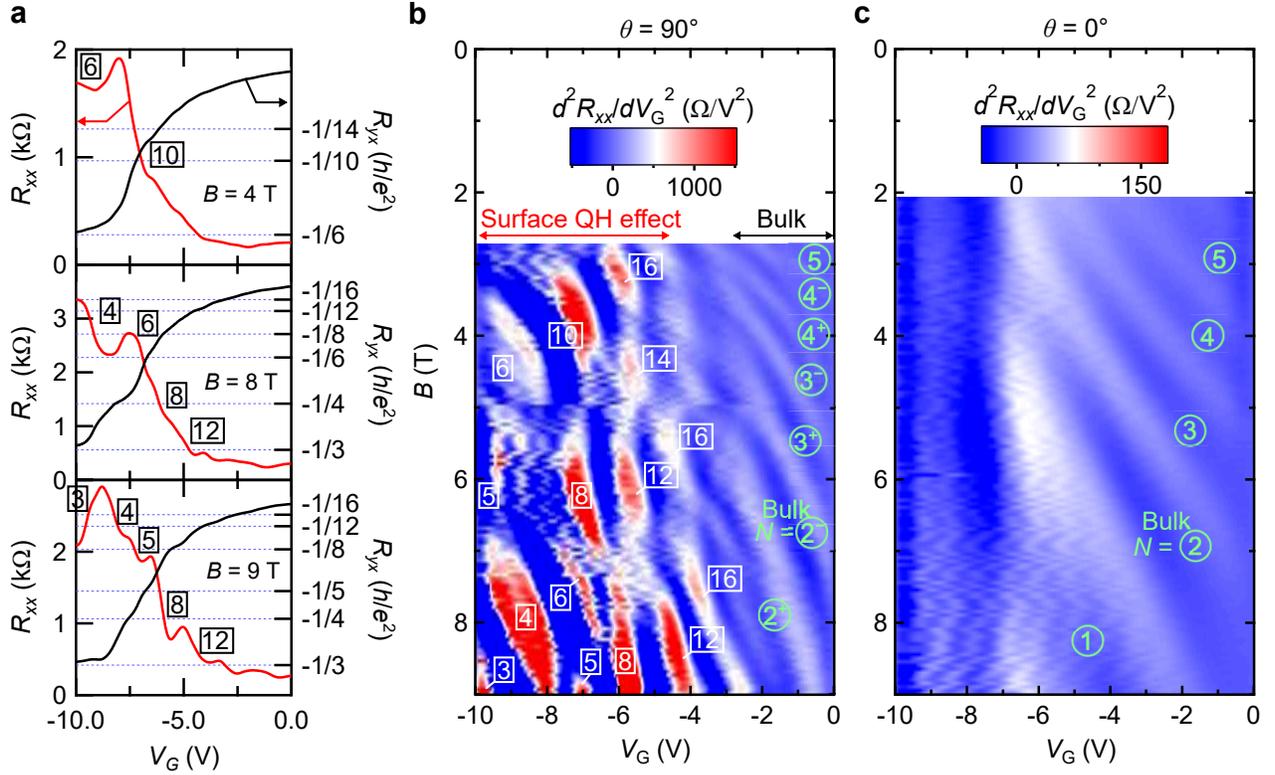}
\caption{
\textbf{Gate-modulation of the surface quantum Hall state}. \textbf{a}, Gate-voltage {\VG} scans of transverse magnetoresistance (MR) at representative fields ($B$ = 4, 8, and 9 T) for a $x = 0.19$ sample with $t$ = 95 nm. Application of a negative bias ({\VG} $< 0$) corresponds to the depletion of electrons in the film. With negatively increasing {\VG}, Weyl orbit QH states develop with quantization in Hall resistance $R_{yx}$ (right axis). \textbf{b},\textbf{c}, Contour mapping of the second derivative of transverse MR and longitudinal MR plotted as a function of {\VG} and $B$. In the transverse MR, the surface QH effect starts to dominate the bulk conduction with reducing the electron density, while only the bulk Landau levels are observed in the longitudinal MR. The QH states are initially characterized by a degeneracy of 4, which is reduced to 2 and finally to 1 with increasing {\VG} ($< 0$) and/or $B$.
}
\end{center}
\end{figure}
\end{document}